# PARTICLE IDENTIFICATION IN GROUND-BASED GAMMA-RAY ASTRONOMY USING CONVOLUTIONAL NEURAL NETWORKS

E.B. Postnikov[1,a], I.V. Bychkov[2,3], J.Y. Dubenskaya[1], O.L. Fedorov[4], Y.A. Kazarina[4], E.E. Korosteleva[1], A.P. Kryukov[1], A.A. Mikhailov[2], M.D. Nguyen[1], S.P. Polyakov[1], A.O. Shigarov[2,3], D.A. Shipilov[4], D.P. Zhurov[4]

[1] *Lomonosov Moscow State University, Skobeltsyn Institute of Nuclear Physics (SINP MSU), 1(2) Leninskie gory, GSP-1, Moscow, 119991, Russian Federation*

[2] *Matrosov Institute for System Dynamics and Control Theory, Siberian Branch of Russian Academy of Sciences, 134 Lermontov st., P.O. Box 292, Irkutsk, 664033, Russia*

[3] *Irkutsk State University, 1 Karl Marx st., Irkutsk, 664003, Russia*

[4] *Applied Physics Institute of Irkutsk State University (API ISU), 20 Bul'var Gagarina, Irkutsk, 664003, Russia*

E-mail: [a] evgeny.post@gmail.com

Modern detectors of cosmic gamma-rays are a special type of imaging telescopes (air Cherenkov telescopes) supplied with cameras with a relatively large number of photomultiplier-based pixels. For example, the camera of the TAIGA-IACT telescope has 560 pixels of hexagonal structure. Images in such cameras can be analysed by deep learning techniques to extract numerous physical and geometrical parameters and/or for incoming particle identification. The most powerful deep learning technique for image analysis, the so-called convolutional neural network (CNN), was implemented in this study. Two open source libraries for machine learning, PyTorch and TensorFlow, were tested as possible software platforms for particle identification in imaging air Cherenkov telescopes. Monte Carlo simulation was performed to analyse images of gamma-rays and background particles (protons) as well as estimate identification accuracy. Further steps of implementation and improvement of this technique are discussed.

# 1. Introduction

Ground-based gamma-ray astronomy studies very energetic radiation of galactic and extragalactic origin by specially designed telescopes, the so-called Imaging Air Cherenkov Telescopes (IACTs) [1]. With this technique, gamma-rays are observed on the ground optically via the Cherenkov light emitted by extensive showers of secondary particles in the air when a very-high-energy gamma-ray strikes the atmosphere. Gamma-rays of such energies contribute only a fraction below one per million to the flux of cosmic rays [2], most of which are protons. Nevertheless, being particles without electric charge they can be extrapolated back to their origin, which makes them the best "messengers" of exotic and extreme processes and physical conditions in the Universe.

That is why particle identification (gamma-ray discrimination against the cosmic-ray background) is an essential part of data analysis for the IACT technique. The study compares two open source machine learning libraries, PyTorch [3] and TensorFlow [4], as software platforms for solving this important problem of data analysis in gamma-ray astronomy. The TAIGA-IACT telescope [5] Monte Carlo simulation is used for this purpose.

The work is a component of the Russian-German astroparticle data life cycle initiative named as Astroparticle.online [6]. The aim of the project is the development of an open science system for collecting, storing, and analysing astroparticle physics data. Astroparticle physics studies elementary particles of astronomical origin including gamma-rays at very high energies. It's a modern interdisciplinary field of research at the intersection of particle physics, astronomy, astrophysics, and cosmology. Currently, the project collaborates with the TAIGA [5] and KASCADE-Grande [7] experiments and invites astroparticle experiments to participate.

# 2. Monte Carlo simulations

The simulation was performed to obtain datasets with the response of a real IACT telescope for two classes of particles to be identified: gamma-rays and background particles (protons). A power-law energy spectrum of particles with the experimental value of the slope was used to generate the datasets. The IACT pointing direction spanned the zenith angle range of 30−40° corresponding to the Crab Nebula observations of TAIGA-IACT [5]. Protons were incident within up to 10° around the fixed pointing direction of the IACT to span the full telescope aperture (the total field of view diameter is 9.6°), whereas gamma-rays were incident within up to 0.05° around the IACT pointing direction, which is the expected IACT pointing and tracking accuracy.

The development of the shower of secondary particles in the atmosphere was simulated with the CORSIKA package [8]. The response of the IACT system was simulated using the OPTICA-TAIGA software developed at JINR [9]. It describes the real TAIGA-IACT setup configuration: 29 constituent mirrors with an area of about 8.5 $m^2$ and a focal length of 4.75 m, and the 560-pixel camera located at the focus, each pixel being a photomultiplier (PMT) collecting light from the mirrors. The telescopic image was formed using a dedicated software developed at SINP MSU taking into account the night sky background fluctuations, PMT characteristics, and triggering and readout procedures of the data acquisition system. A final image to be analysed is obtained by subtracting the average night sky background value (pedestal) from the PMT pulse amplitudes after readout in each PMT.

# 3. Conventional methods

### 3.1. Image cleaning

The procedure of an IACT image reconstruction above the night sky background is called image cleaning; it's intended to remove images (and parts of images) produced by the night sky background fluctuations instead of a shower of secondary particles. The conventional procedure is two-parametric: it excludes from subsequent analysis all image pixels except the "core pixels", i.e. those with the amplitude above a "core threshold" and at least one neighbour pixel above a "neighbour threshold", and the neighbour pixels themselves. If the image contains too few pixels after cleaning (for example, 2 or less), the entire image is excluded from the analysis.

Examples of simulated images before and after cleaning with a low threshold (core threshold is 6 photoelectrons (p.e.) and neighbour threshold 3 p.e., whereas an average night sky background amplitude is 2.6 p.e.) are given in Figure 1. The difference between the original images and those after cleaning is not sufficient because of the low cleaning threshold.

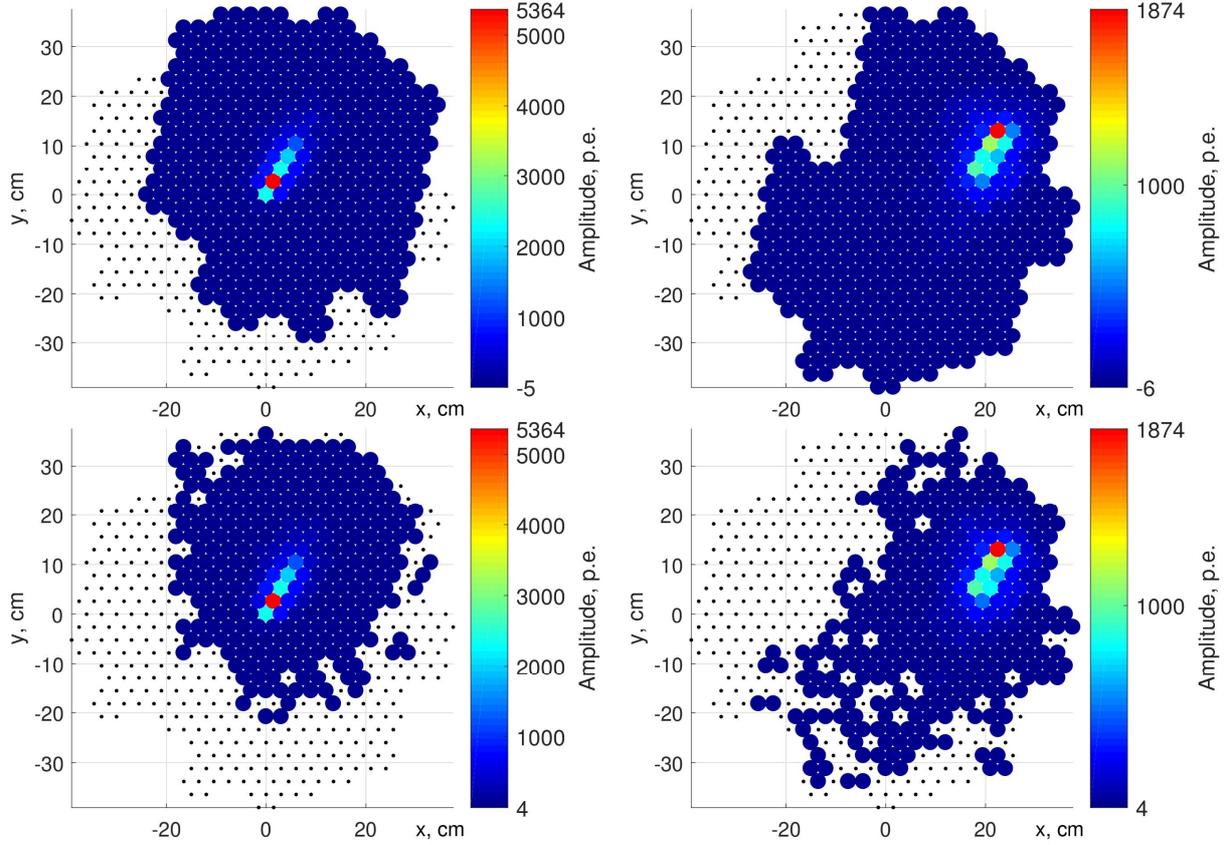

Figure 1. Gamma-ray (left panel) and proton (right panel) images before image cleaning (top panel) and after image cleaning with a low threshold (bottom panel)

Two different studies were performed: in the first one, deep learning algorithms were trained on images without cleaning, whereas in the second one, they were trained on images after cleaning. However, for the reference technique, a test sample was first subjected to the image cleaning procedure in any case. No training sample was needed for the reference technique.

**3.2. Hillas parameters**

The conventional way to solve the particle identification problem in IACT data analysis is a parametrization of the image using empirical variables named after Hillas [10]: length, width, angle, and their derivatives. These variables and their combinations can serve as discriminators between gamma-ray and proton for two reasons. First, on average gamma-ray images are narrower than proton ones. Second, they are oriented to the point of gamma-ray source projection onto the image plane (for the point-source observations, such as the Crab Nebula, it's the centre of the camera's field of view), whereas proton image orientation is completely random.

**3.3. Reference method**

For comparison with the deep learning approach, two Hillas parameters most informative for discrimination between gamma-rays and protons were calculated: the image width $w$ and angle $\alpha$ between the image orientation and the direction to the camera centre. These two parameters allow discriminating gamma-ray images (as those with small $w$ and $\alpha$) against proton background. As the reference technique, a set of two consecutive cuts (on $w$ and $\alpha$) was chosen with the optimal cut values found as maximizing the $Q$ value (next section) on the learning samples of simulated gamma-rays and protons under the condition that the number of incorrectly identified gamma-rays is below 50%.

### 3.4. Quality criterion

The quality criterion for particle identification was the selection quality factor or Q factor. This factor indicates an improvement in the significance of a gamma-ray signal above background compared with the significance before selection [11]. For the Poisson distribution (that is for a large number of events), the selection quality factor is:

$$Q = \varepsilon_\gamma / \sqrt{\varepsilon_{bckgr}} , \qquad (1)$$

where $\varepsilon_\gamma$ and $\varepsilon_{bckgr}$ are relative numbers of gamma-ray events and background events after selection. In this study, protons are considered as background, and gamma-rays were selected above background.

## 4. Convolutional Neural Network (CNN)

Under the deep learning approach, a CNN was used because it's well-adapted to classify images; it was also the choice for all deep learning applications to the IACT technique [12, 13]. The advantage of CNN is a fully automatic algorithm, including automatic extraction of image features instead of Hillas parameters. It is implemented in free software packages: PyTorch [3] and TensorFlow [4]; both of them were tested in this study. In contrast to work [12] studying square pixels, hexagonal shape and arrangement of pixels of TAIGA-IACT have not been fully taken into account yet, but only an approximation of the regular square grid using oblique coordinates with angle 60° was used.

Training datasets contained gamma-ray and proton images (Monte Carlo of TAIGA-IACT, energy distributions in the range 2−60 and 3−100 TeV respectively with the spectral slope -2.6). Test datasets (different from training ones) of gamma-ray and proton images in random proportion (blind analysis) were classified by each of the packages: TensorFlow and PyTorch. Various networks with different parameters were tested to find the one maximizing the Q factor.

## 5. Results

Automatic image feature extraction by CNN made us first trying gamma-ray identification using Monte Carlo images without image cleaning (section 3.1) at all. For that purpose training and test samples with neither cleaning nor preselection were analysed by both PyTorch and TensorFlow.

Next effort was made with training and test samples after image cleaning. In case the image had less than 3 pixels after cleaning, it was removed from the analysis before calculating the quality factor. Quality factor (1) values obtained by the best CNN configuration among all the trained networks are assembled in Table 1 together with the quality factor for the reference technique.

Table 1. The quality factor for gamma/proton separation

|  | CNN (PyTorch) | CNN (TensorFlow) | Reference (Hillas analysis) |
|---|---|---|---|
| Without image cleaning | 1.74 | 1.48 | 1.76 |
| With image cleaning | 2.55 | 2.99 | 1.70 |

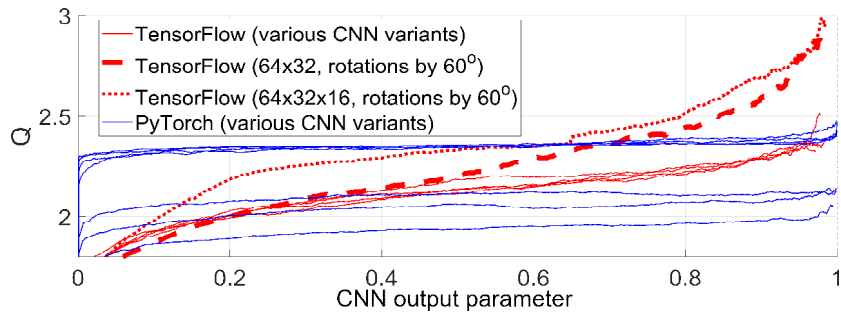

Figure 2. Quality factor vs CNN output parameter (a scalar parameter between 0 and 1 characterizing image similarity to gamma-ray or proton)

Comparison of different CNN versions for both software packages is illustrated in Figure 2. Overall PyTorch had more stable results in a wide range of CNN output parameter values. However, significant improvement was obtained with the TensorFlow CNN version trained by a modified training sample, which contained both original images and those rotated 60° (symmetry angle of hexagonal structure) and thereby consisted of ~180000 events instead of ~30000. Therefore, the performance of different software packages was approximately equal, and the training sample size was crucial for the performance. The PyTorch CNN has not been tested with the modified training sample.

## 6. Acknowledgement

The work was supported by the Russian Science Foundation, grant #18-41-06003.

## 7. Conclusion

Both CNN packages were tested and showed approximately equal results. Two algorithms were found to give a significant improvement in CNN separation quality: image cleaning, even with a low threshold, and additional image rotation in the training sample, which allows increasing sample size. The results can be further improved by taking into account the hexagonal pixel shape and increasing training sample size by one order of magnitude.